\documentclass[aps,onecolumn,showpacs,superscriptaddress,notitlepage]{revtex4}
\usepackage{latexsym,amssymb}
\usepackage{graphicx}
\usepackage{amsmath}
\usepackage{epstopdf}
\usepackage{placeins}
\usepackage{ulem}

\begin{document}
\title{Localization properties of one-dimensional speckle potentials in a box}

\author{J. Giacomelli}
\affiliation{Sace, Piazza Poli 37/42, 00187 Roma, Italy}

\begin{abstract}
We investigate the localization properties of the single particle spectrum of a one-dimensional speckle potential in a box. We consider both the repulsive and the attractive cases. The system is controlled by two parameters: the size of the box and a rescaled potential intensity. The latter is a function of the particle mass, the correlation length and the average intensity of the field. Depending on both these parameters values and the considered energy level, the eigenstates exhibit different regimes of localization. In order to identify the regimes for the excited states, we use a technique developed in this work. Depending on the chosen parameters values, we find that it is possible not observing any effective mobility edge nor delocalization of the eigenstates due to the finite size of the system.
\end{abstract}
{\let\newpage\relax\maketitle}
\section{Introduction}
\label{intro}
Random potentials created by laser speckles are routinely employed in experiments with ultracold atoms to investigate the behaviour of disordered systems \cite{fallani2008,shapiro2012, sp2010, gmodugno2010}.
One-dimen\-sional (1D) speckle potentials in particular have been the object of an intensive study in recent years. From both the experimental and the theoretical side, many interesting features have been addressed. These include classical localization and fragmentation effects, frequency shifts and damping of collective excitations, inhibition of transport properties, Anderson localization and related phenomena \cite{damski2003,lye2005,fort2005,modugno2006,clement2006,lugan2007,shapiro2007,sanchezpalencia2007,billy2008,lugan2009,piraud2011}.

The theoretical investigation on the 1D speckle potential and, more in general, on the 1D disordered systems focuses on the infinitely extended case and nowadays a clear framework has been outlined. In 1D disordered systems, all states are exponentially (aka Anderson) localized, regardless to the weakness of the random potential \cite{anderson1958}. The quantity that is commonly used to describe this phenomenon is the eigenstates inverse localization length (or Lyapunov exponent), which has been shown to be strongly dependent from the shape of the considered potential. In the Born approximation regime, the Lyapunov exponent as a function of energy results to be simply proportional to the Fourier transform of the potential autocorrelation function  \cite{sanchezpalencia2007,izrailev1999}. Therefore, if the power spectrum of a given disordered potential has a finite support (i.e. if the potential self$-$correlation is long ranged), Lyapunov exponent goes to zero beyond the critical energy $E_C$ associated with the highest available Fourier component. This fact defines the so called effective mobility edge. The incongruity between the prediction above and the presence of localization at every energy level can be solved by considering the perturbative expansion of the Lyapunov exponent beyond the Born approximation (in \cite{lugan2009} this is proven for the 1D speckle potentials). In fact, the localization length remains finite also beyond $E_C$, due to the contribution of the $n>2$ sites correlators, though showing a kink while crossing the energy threshold. Since all the eigenstates are localized, the effective mobility edge is not, in any sense, an Anderson transition and so the 1D disordered systems do not exhibit any multifractal behaviour at $E_C$ \cite{evers2008}. 

Other interesting results have been obtained for the 1D speckle potentials. Regarding the uniform case, the low energy behaviour has been first considered in \cite{lugan2007}, where it has been shown that density of states in a repulsive speckle potential is characterized
by a Lifshitz tail \cite{lifshitz1964,lifshits1988}. As for both attractive and repulsive speckles, this has been discussed more thoroughly in \cite{falco2010}, where it has also been shown that three different regimes of speckle intensities $s$ can been identified at low energy - semiclassical ($s\gg1$), intermediate, and  quantum ($|s|\le1$). The presence of an effective mobility edge have been  observed in the Anderson localization experiment \cite{billy2008}. Furthermore, the existence of both \textit{extended} (compared with the finite length scale of the system) and localized states in the presence of an inhomogeneous confinement has been discussed in \cite{pezze2011}.

Despite this intense research activity, the single particle spectrum properties for a speckle potential have been only partially addressed, even in the 1D case. In fact, the framework summarized above has been developed considering only infinitely extended systems. The finite sized system is usually employed just as a numerical tool to verify the theoretical predictions, provided that it is large enough to be a good approximation of the infinitely extended case. What is lacking so far is a detailed numerical analysis of how the eigenstates density profile of the finite sized speckle system is related to the main properties of the potential depending on the chosen energy level (i.e.: spatial extension, intensity, blue or red detuning). More specifically, we want to investigate the 1D speckle systems with an intermediate size between the \textit{large} ones, reproducing the infinitely extended case behaviour, and the \textit{short} ones, where the localization length is greater than the size and so the eigenstates are extended. For these systems the correlators cannot be properly defined beyond the system length scale and a similar consideration holds for the definition of the Lyapunov exponent. To the best of our knowledge, there is not a clear picture that describes to what extent the finite size affects predicted behaviours such as the effective mobility edge. 

In this article we consider a 1D speckle potential delimited by infinite walls - a sort of \textit{disorder in a box} - and we discuss the localization properties of its eigenstates as the speckle intensity $s$ and the size of the system $L$ changes. As discussed in \cite{meyrath2005}, the box-like barriers 
can be achieved experimentally by using two laser beams that propagate perpendicular to the speckle direction,  allowing for the investigation of finite size effects in a textbook case.

We study both the ground state and the excited states, with the aim of characterizing the system localization properties while varying the parameters $s$ (amplitude of the potential), $L$ (size of the system) and the energy level $E$. When possible, we extrapolate the system behaviour in the limit $L\rightarrow\infty$.

We observe that the center of mass of some of the eigenstates shifts stochastically while varying $s$, this depending on the single realization of the speckle potential. A numerical measure of the probability of observing this phenomenon is compared with a theoretical framework developed in this work. We are then able to separate the contribution due to the boundary effects from the one depending only on the shape of the disordered potential. We express the latter as a function of the participation ratio.

We address the role played by the finite size of the system $L$ and we compare our numerical results with the theory. Properly choosing $L$ and $s$, we observe a regime which is different from the predicted one for $L\rightarrow\infty$. In this regime, eigenstates are still localized at $E\geq E_C$ ($L$ is big enough to make the boundaries role not to be predominant). However, the effective mobility edge is completely absent. As $L$ gets bigger, we observe the approach of the system to the predicted behaviour.

Although the considered system is not able to show a true multifractal behaviour, due to the absence of any Anderson transition, we use some traditional tools of multifractal analysis to define a functional able to distinguish localized eigenstates from extended ones and from the ones belonging to the crossover region of the spectrum. We verify that this observable is more effective than traditional ones and so we use it to characterize the three regions of the spectrum while varying $s$.

The article is organized as follows. In Sec.~\ref{intrononint} we define both the system we consider and the main tools we use to characterize its eigenstates.
In Sec.~\ref{secgs} we discuss the properties of the ground state and how its appearence depends on $s$ and $L$.
The density profile of the excited states is considered in Sec.~\ref{seces}. Our results are summarized in~Sec.~\ref{secend}.

\section{Model and methods}
\label{intrononint}
Let us consider a single particle in a 1D speckle potential $V_{s}(x)=V_0 v(x/\xi)$, with  intensity $V_0 = \left\langle V_s\right\rangle$ and autocorrelation length $\xi$ \cite{goodman2005,modugno2006}. The probability distribution of $v(x)$ is $e^{-v}$. Moreover it holds that
\begin{equation}
\langle v(x)v(x+\Delta)\rangle = 1+sinc^2\left(\frac{\Delta}{\xi}\right)\nonumber
\end{equation}
Optical speckle is obtained by transmission of a laser
beam through a medium with a random phase profile, such as
a ground glass plate. The resulting complex electric field
is a sum of independent random variables and forms a Gaussian process. Atoms experience a random potential proportional to the intensity of the field. $V_0$ can be both positive or negative, the potential resulting in a series of barriers or wells, that in the following will be referred as \textit{repulsive} (or blue) and \textit{attractive} (or red), respectively. The different features of the system in the red and the blue cases arise from the different distributions of speckle intensity maxima and minima. In fact, the maxima distribution is approximately proportional to $\sqrt{I_{max}}\exp{(-I_{max})}$ while the minima distribution diverges in zero \cite{falco2010}. 
The autocorrelation length $\xi$ represents a natural  scale for the system, and $E_{\xi}=\hbar^2/2m\xi^2$ the corresponding energy scale. The Hamiltonian of the system can be written in a dimensionless form as
\begin{equation}
H = -\frac{d^2}{dx^2} + s v(x),
\label{hamiltonian}
\end{equation}
where the rescaled intensity $s={2m\xi^2V_0}/{\hbar^2}$ represents the only parameter of the system. In case of a finite size system also the size $L$ becomes relevant; its effects will be discussed hereinafter.

The speckle pattern can be generated numerically as discussed in \cite{modugno2006} (and references therein).
Then, the spectrum can be obtained by solving the stationary Schr\"odinger equation $H\psi=E\psi$, by mapping $H$ on a grid with vanishing boundary conditions, corresponding to infinite walls.

For a fixed system size $L$, the localization behaviour of the eigenstates in the different intensity regimes of speckle potential have been outlined in \cite{falco2010}. 
The characteristic extension and the localization properties of the eigenstates depend considerably on $s$, and one can identify three different regimes,
ranging from the semiclassical limit, for $|s|\gg1$, down to
the quantum regime, when $|s|\le1$ (with an intermediate regime in-between).
In the case of attractive speckles, and for shallow potentials (quantum regime), the eigenstates 
extend over several potential wells. The lowest lying eigenstates localise around deep wells but the ground state is not necessarily localized in the deepest well as the width of the well plays also a crucial role. When $\vert s\vert$ is increased, the eigenstate width shrinks and their position may change. 
Eventually, for large enough $\vert s\vert$ (semiclassical regime), the eigenstates tend to stack up as bound states inside isolated wells, with the ground state being the lowest eigenstate of the deepest well. 
A similar scenario holds for  repulsive speckles, in this case the localization taking place in between barriers. 

Here we discuss it more thoroughly, characterizing the spectrum as a function of $s$, and discussing its dependence on $L$. In particular we consider a system length typically in the range $L/\xi\in[200,1000]$ (up to $L=7000$ in certain cases), and average over $50$ to $500$ speckle realizations (see \ref{appnum} for further details).  

In order to characterize the localization properties of the eigenfunctions we use the following quantities.  

\begin{enumerate}
\item 
The participation ratio 
\begin{equation}
\label{iprdef}  
PR\left\lbrack\psi\right\rbrack =\frac{1}{L\int_{L} dx \left\vert\psi(x)\right\vert^4}
\end{equation}  
that measures the  relative extent occupied by the state $\psi$; it decreases as
 $\psi$ gets more and more localized.  
 \item 
The localization length $\ell_{loc}$, that characterizes the exponential decay of the tails of Anderson localized states, $|\psi(x)|^{2}= C\exp\{-2\vert x-x_{0}\vert/\ell_{loc}\}$, with $x_{0}\equiv\textrm{argmax}\left\{\left\vert\psi (x)\right\vert^2\right\}$ being the localization center and $C$ such that $\int_0^L\vert\psi(x)\vert^2dx=1$.
\newline Since $L$ is finite, the exponential decay occurs over two finite domains $D_1\subseteq [0,x_{0}]$  and $D_2\subseteq [x_{0},L]$. Moreover, the exponential density profile is locally perturbated by boundary effects and fluctuations of the disorder. Hence we define and measure $\ell_{loc}$ over a tail as
\begin{equation}
\ell_{loc} =-\frac{2}{a},\quad (a,b)=\underset{a',b'}{\textrm{argmin}}\int_{D_i}dx\left(\ln|\psi(x)|^{2}-a'x-b'\right)^2
\end{equation} 
This definition allows us to measure $\ell_{loc}$ even when the exponential decay is slightly warped by boundaries or local disorder. See \ref{appnum} for further details on this quantity and the next two. 
\item
The average curvature of the wave function tails, defined through the following functional
\begin{equation}
\label{eqeta}
\eta(s,L) = \frac{1}{2}\sum_{i=1}^{2}\frac{1}{L_{i}}\left\vert\int_{D_i}\frac{d^2 \ln |\psi(x)|^{2}}{dx^2}dx\right\vert,
\end{equation} 
where $L_{i}=\int_{D_i}dx$ is the extent of left and right tails (over the ground noise).
$\eta$ vanishes for exponentially localized eigenstates, and is positive definite otherwise. In fact, extended eigenstates show a positive $\eta$ due to the boundaries of the box, and states being more than exponentially localized show a positive $\eta$ too, due to the local shape of disorder. However, we especially use $\eta$ in order to measure the local effect of the disorder on the tails.  
\item
The \textit{localization volume} $D_{loc}=L_1+L_2$, defined as the length of the interval over which the exponential localization occurs. We use the convention $D_{loc}=L$ in the limit of a fully delocalized state, despite the lack of any exponential decay in this case.
\item
The functional $\sigma^2_{\omega}\left[\cdot\right]$ defined as
\begin{equation}
\label{sodef}
\sigma_{\omega}^2\left[\psi\right]=\sum_q\sigma^2_{\Lambda}\left[\frac{\partial}{\partial\ln\Lambda}\ln\sum_{k=1}^N\left(\int_{(k-1)\Lambda}^{k\Lambda}dx\left\vert\psi(x)\right\vert^2\right)^q\right]
\end{equation} 
where $N=\frac{\Lambda}{L}$ and $\sigma^2_{\Lambda}\left[f(\Lambda,\dots)\right]$ is the variance of $f$ as a function of $\Lambda$. We introduce $\sigma_{\omega}^2\left[\psi\right]$  in order to characterize the eigenstates of the system by relating their localized/extended behaviour with their grade of self-similarity. The derivation of this definition can be found in  \ref{app_sodef}.
\end{enumerate}
The first two are widely used in the literature, while the others are defined here in order to show specific features of this system that more commonly employed functionals are not effective in measuring.

In the next section we start by considering the properties of the ground state as a function of $s$, discussing their dependence on the finite size of the system. Then, in the following section we will consider the behaviour of the full spectrum. 

\section{The ground state}
\label{secgs}

\begin{figure}[h]
\centerline{\includegraphics[width=0.99\columnwidth]{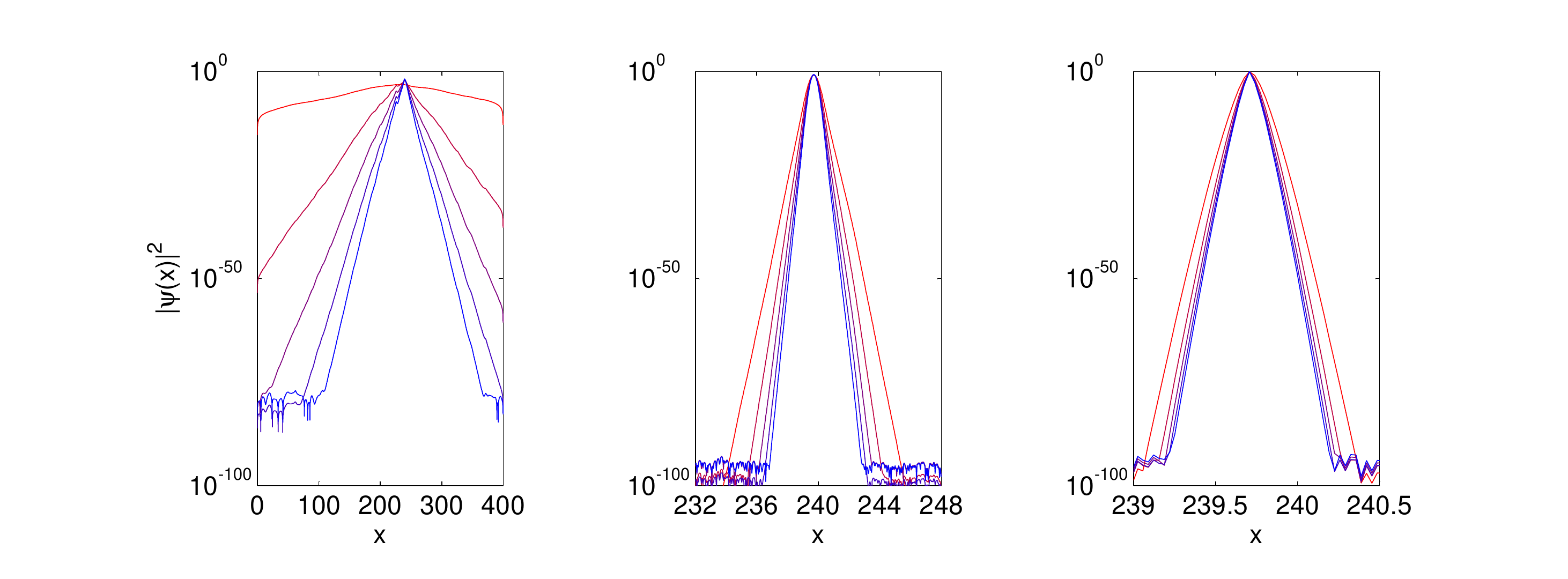}}
\caption{Behaviour of the density $\left\vert\psi(x)\right\vert^2$ of the ground state for a single realization of red-detuned speckle potential $(L=400)$. We have considered the following intervals of $\vert s\vert$, respectively from the left to the right: $ 2\cdot10^{-3}<\vert s\vert<2\cdot10^{-1}$ (left);  $ 4<\vert s\vert<40$ (center);  $ 10^{5}<\vert s\vert<10^{6}$ (right). We plot the considered values of $\vert s\vert$ in increasing order from red to blue line.  }
\label{figesempio}
\end{figure}

For a fixed system size $L$, the localization behaviour of the eigenstates in the different intensity regimes of speckle potential have been outlined in \cite{falco2010}. Here we discuss it more thoroughly.
We start by considering the density distribution of the ground state for a single realization of length $L=400$, as a function of $s$, as shown in Fig. \ref{figesempio}. As an example, we discuss the case of a red detuned speckle realization; a similar behaviour is observed for the blue detuned case as well. In the attractive case ($s<0$), the ground state is defined due to the finite size of the system. Otherwise, at $L\rightarrow\infty$ we have $\left\langle\min[v(x)]\right\rangle\rightarrow-\infty$ and so the spectrum is not inferiorly limited. 

\FloatBarrier

According to our simulations, some of the behaviours observed in Fig. \ref{figesempio} have a general validity: 

$(i)$ For too small values of $|s|$ the localization behaviour of the ground state is affected by the boundaries ($\vert s \vert\simeq 10^{-3}$). In this case the tails decay exponentially, but the finite size of the system limits the localization length $\ell_{loc}$ to values of the order of the system size $L$. 

$(ii)$ In the opposite limit, for large values of $|s|$ the ground state is bounded in a single well (red speckles) or between two adjacent barriers (blue speckles). In this case the density profile of the tails tends to be Gaussian.

$(iii)$ For intermediate values of $|s|$ the behaviour of the tails continuously connects the two extreme regimes. 

\begin{figure}[h]
\centerline{\includegraphics[width=0.95\columnwidth]{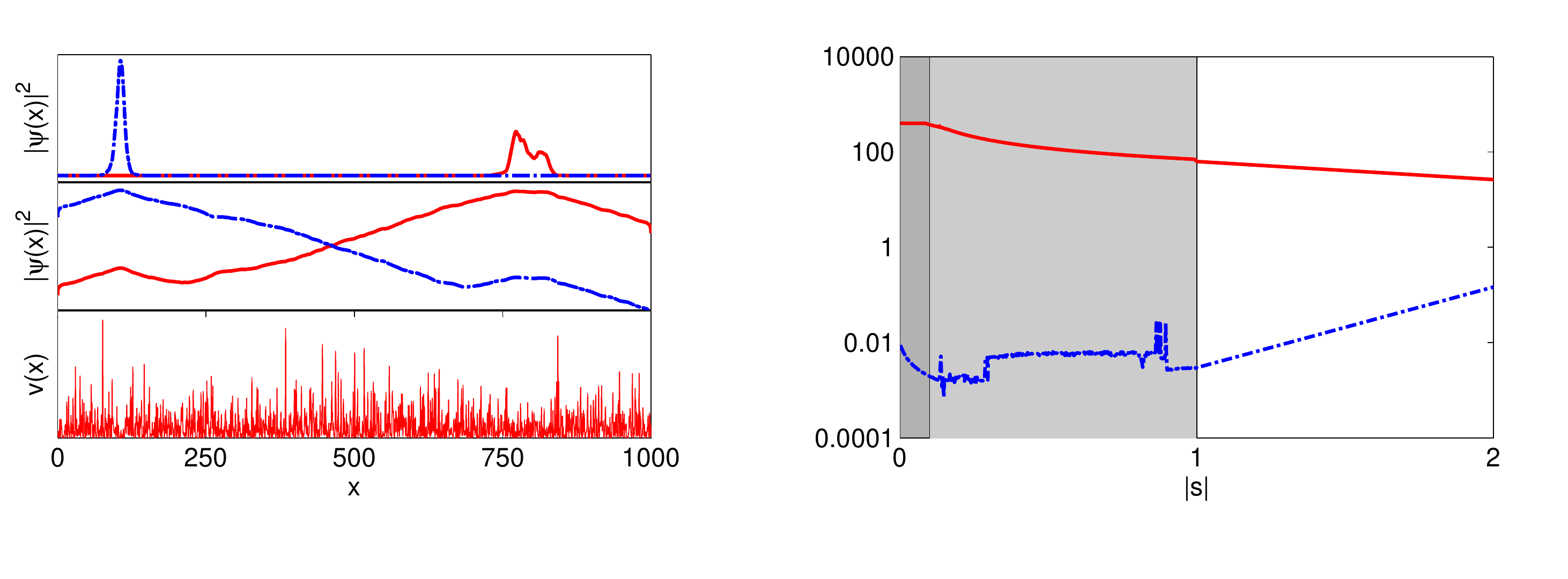}}
\caption{$(left)$ An example of the change of localization position for increasing $\vert s\vert$. In the example we considered $L=1000$ and $s\simeq0.1$. The two ground states are obtained with a shift $\Delta s\simeq 10^{-7}$. $(right)$ $\eta(s)$ (blue dashed curve) and $D_{loc}(s)$ (red solid curve) for the ground state of the speckle realization considered in fig. \ref{figesempio}.}
\label{figesempioetaext}
\label{figesempioclp}
\end{figure}

$(iv)$ For some speckle realization, the localization center of the ground state may change as a function of $s$, see e.g. Fig. \ref{figesempioclp}. This effect will be discussed in \S \ref{clppar}.

Two of the quantities defined in \S \ref{intrononint} - $\eta(s)$ and $D_{loc}(s)$ - allow for measuring the interval of $\vert s\vert$ where there is a boundary effect (dark grey area), as shown in Fig. \ref{figesempioetaext} for the same speckle realization used for  Fig. \ref{figesempio}. In fact $D_{loc}(s)$ remains constant and $\eta(s)$ decreases until $\vert s\vert\gtrsim -10^{-2}$, when the tails detach from the boundaries of the system. The decrease of $\eta(s)$ is linked to the local curvature of the tails when they are too near to the boundaries and, for even smaller $\vert s \vert$ values, to the fact that the ground state cannot present a completely exponential localization.
When the boundary effects are over (Fig. \ref{figesempioetaext}, light grey area), there is an interval of $\vert s\vert$ where $\eta(s)\simeq 0$: a ``pure'' Anderson localization occurs. By increasing  $\vert s\vert$ we observe that the average curvature of the ground state begins to grow up, due to the increasing importance of the deep fluctuation of the potential where the speckle is localized. We have not observed any clear dependence of the amplitude of the region where a ``pure'' Anderson localization occurs from $s$ and $L$, since there is a very strong dependence of $\eta(s)$ from the shape of the single speckle potential, until it starts to constantly grow up. In fact, the most of the considered speckle realizations show a noisy behaviour of $\eta(s)$ for small $\vert s\vert$ values, due to the presence of big local fluctuations of the potential and to random changes of the localization position.

However, all of the considered speckle realizations show an abrupt rise of $\eta(s)$ starting from a $s$ value between $\vert s\vert\simeq 1$ and $\vert s\vert\simeq 3$ (considering $100\leq L\leq500$). Figure \ref{figesempioclp} (right panel) shows an example of $\eta(s)$ that begins rising at $\vert s\vert\simeq 1$.

\FloatBarrier
\subsection{Localization type}
\label{parloctype}

\begin{figure}[h]
\centerline{\includegraphics[width=1.05\columnwidth]{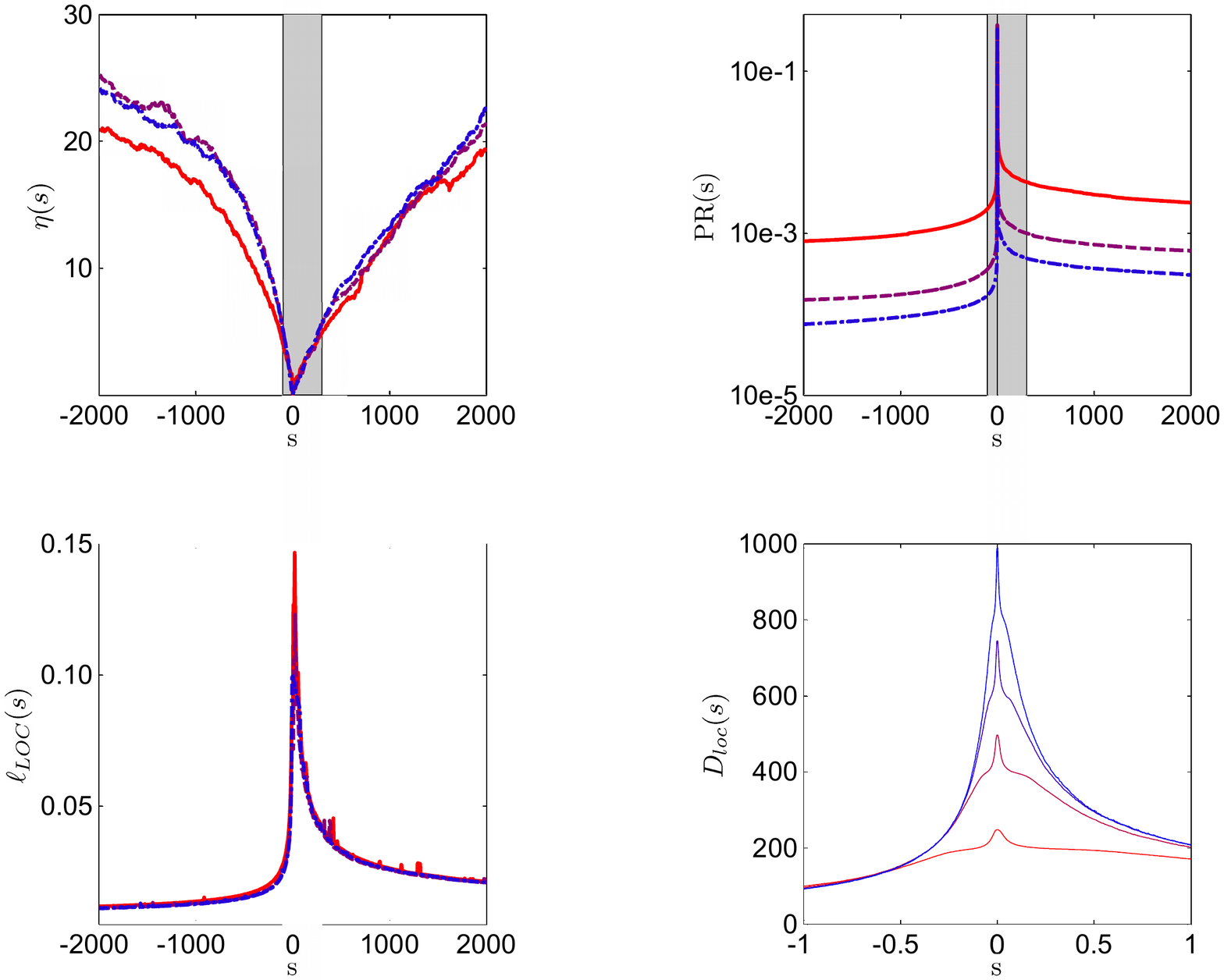}}
\caption{The functionals $\eta$ (left top panel), $PR$ (right top panel) and $\ell_{LOC}$ (left bottom panel) for the ground state as a function of $s$, increasing the size $L$ of the system from $L=100$ (red line) to $L=500$ (purple line) and $L=1000$ (blue line). For each choice of $s,L$ considered, the plotted value is averaged over $50$ realizations of the speckle potential. The functional $D_{loc}$ (right bottom panel) for the ground state as a function of $s$, increasing the size $L$ of the system from $L=250$ (orange line) to $L=1000$ (blue line). For each choice of $s,L$ considered, the plotted value of $D_{loc}(s,L)$ is averaged over 50 realizations of the speckle potential.}
\label{fig3oss}
\label{figext}
\end{figure}
We discuss here how the localization properties of the states - either exponentially localized by the full random profile of the potential or bound states of single fluctuations (in a single well or between two barriers) - depend on $s$ and $L$. 

In the paragraph above we have discussed the localization process of the ground state density $\vert\psi(x)\vert^2$ at increasing $\vert s\vert$. Fig. \ref{fig3oss}  describes the same process averaged over $50$ realizations of the speckle potential. We can observe that there is no transition from extended to localized states for both the attractive and the repulsive cases. In spite of this, both $PR(s)$ and $\ell_{LOC}(s)$ exhibit a ``knee'' at increasing $\vert s\vert$. This knee separates the region where the localizing tails are influenced by the whole extension of the system (grey area) from the one where the leading contribution to the shape of $\vert\psi(x)\vert^2$ is given by the local configuration of the potential around the localization point (white areas). 

Though the knees are not exactly coincident for $PR(s,L)$ and $\ell_{LOC}(s,L)$, their positions on $s$ axis are comparable. The grey area extends conventionally from $s\simeq-100$ to $s\simeq 300$: in fact the localization process is faster for $s<0$, according with the intuitive fact that a deeper well is more effective than two higher next barrier, in containing the localized ground state at increasing $\vert s\vert$. Moreover, the dependence of $PR(s,L)$ and $\ell_{LOC}(s,L)$ on $L$ is negligible (though present). As we will discuss in \S \ref{me_intro}, we will see that this holds only for the ground state case, since $L$ plays a more relevant role for the degree of localization of the eigenstates while increasing $E$.

The curvature $\eta(s,L)$ is not less sensible to $s$ variations in the white areas than in the grey one. This is not surprising: even if an eigenfunction is so localized to be bounded in a single well, it goes on increasing its curvature as the depth of the well dilates. Moreover, $\eta(s,L)$ exhibits a non negligible dependence from $L$ at $\vert s\vert\gg 1$ values. This dependence from the extension of the system is related to a local effect: in fact, a greater $L$ implies a bigger number of wells (couple of barriers) and, consequently, a deeper global minimum (higher couple of next maxima) on average, where the ground state can be bounded more effectively, that leads to a greater $\eta(s,L)$ value. The role of $L$ is more relevant for $s<0$, consistently with the greater binding effectiveness of the wells than the one of the barries, as stated above. 

In  Fig. \ref{figext} (right bottom) we can observe the behaviour of $D_{loc}(s,L)$ averaged over $50$ realizations of the speckle potential. As stated before, $D_{loc}(\vert s\vert\ll 1, L)\simeq L$. However, when $\vert s\vert \gtrsim 1$ the dependence from $L$ disappear for both $s\lessgtr0$.

\subsection{Variation of the localization position}
\label{clppar}
\begin{figure}[h]
\centerline{\includegraphics[width=0.8\columnwidth]{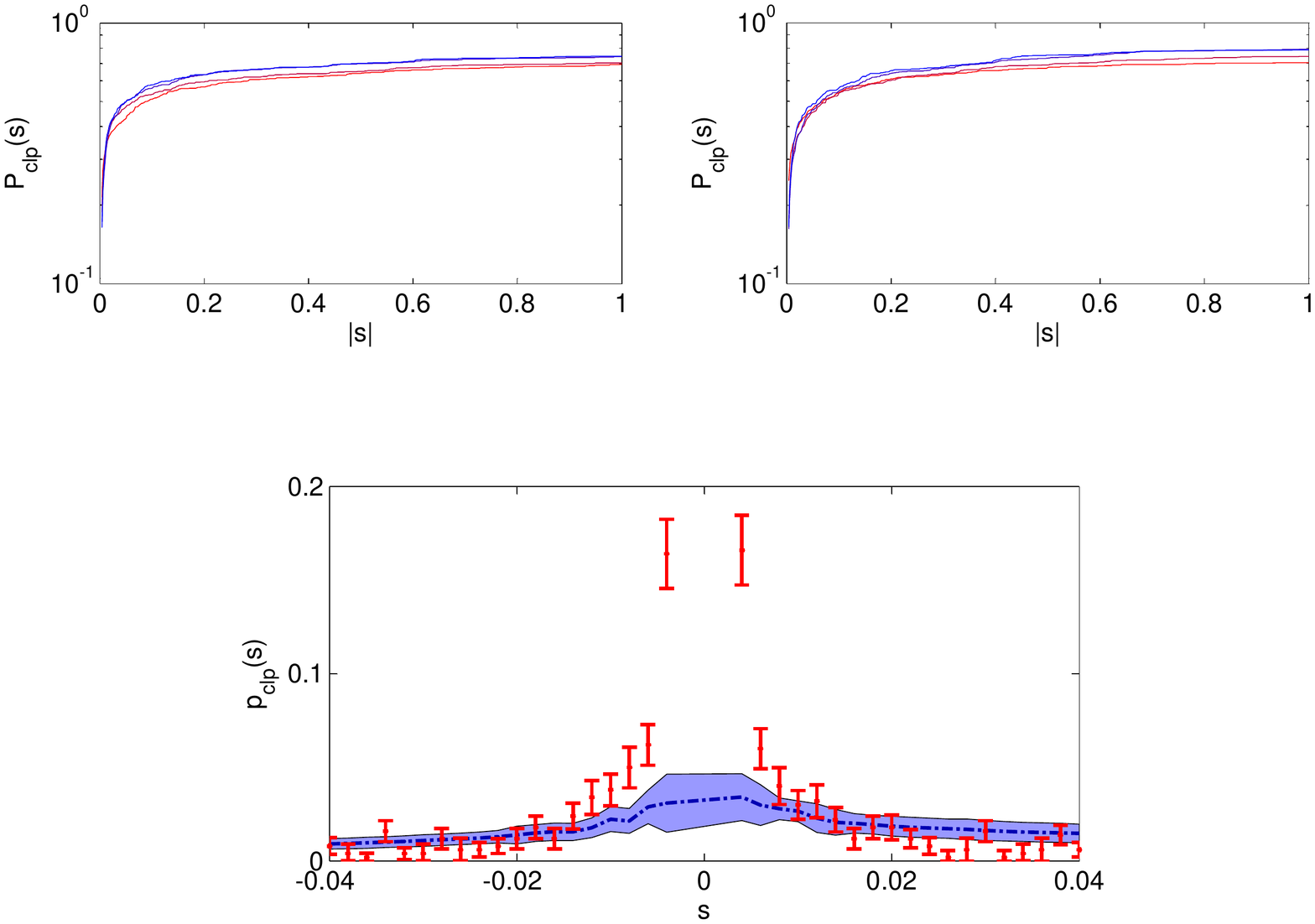}}
\caption{(Top panels) Probability distribution of observing one or more  \textit{change of localization point} 
event for $\vert s'\vert$ lesser than or equal to a given $\vert s\vert$.  The function is obtained simulating the behaviour of the ground state over $500$ speckle realizations as a 
function of $\vert  s\vert$, for both red detuned (left top panel) and blue detuned (right top panel) speckle potential. (Bottom panel) Probability density function for the occurrence of a CLP event. Red dots are computed from the numerical results plotted in figure \ref{figclpcum} for $L=1000$; the blue dashed line is obtained from equation (\ref{eqclpdistr}). The error bars are obtained from the sets of speckle realizations used to compute $p_{clp}$ (red) and $PR$ (blue).}
\label{figclpcum}
\label{figclpthxp}
\end{figure}

In the following we analyze the phenomenon observed in figure \ref{figesempioclp}. The change of localization position (CLP in the following) originates from the possibility that two next eigenstates exchange their position in the energy spectrum. This fact depends on the shape of the single realization of the potential and we found speckles without CLP events on all the tested interval of $s$, as well as speckles which shows this behaviour one or more times.\newline\indent
Given a set of speckle realizations and considering for each potential only the first CLP event observed for the ground state $-$ where possible $-$ while increasing $|s|$, we build the cumulative density function $ P_{clp}(s,L)$, which is a feature of the whole speckle disorder distribution. The result is shown in figure \ref{figclpcum} (Top panels).
With the exception of a small interval of $|s|\simeq 0$, where the boundary effects play a major role in the energy level of the ground state, the probability density function $p_{clp}$ 
of observing a CLP event can be linked with the participation ratio $PR(s,L)$ in a closed form.

\FloatBarrier

Neglecting the local effects of both the disorder and the boundaries, we have the following proxy relation  
\begin{equation}
\label{eqclpdistr}
p_{clp}\left(s,L\right)\simeq\sqrt{\frac{L}{\xi}\frac{PR(s,L)}{2\pi}},
\end{equation} 
This result is obtained in \ref{app_clp}. In figure \ref{figclpthxp} we can observe the crossover between two regions: the one where the boundary effects prevail $(s\simeq 0)$ and the other (large enough $|s|$) where the simulated value of 
$p_{clp}$ shows a good level of agreement with the prediction of equation (\ref{eqclpdistr}).

\section{Excited states}
\label{seces}

\begin{figure}[ht]
\includegraphics[width=1.05\columnwidth]{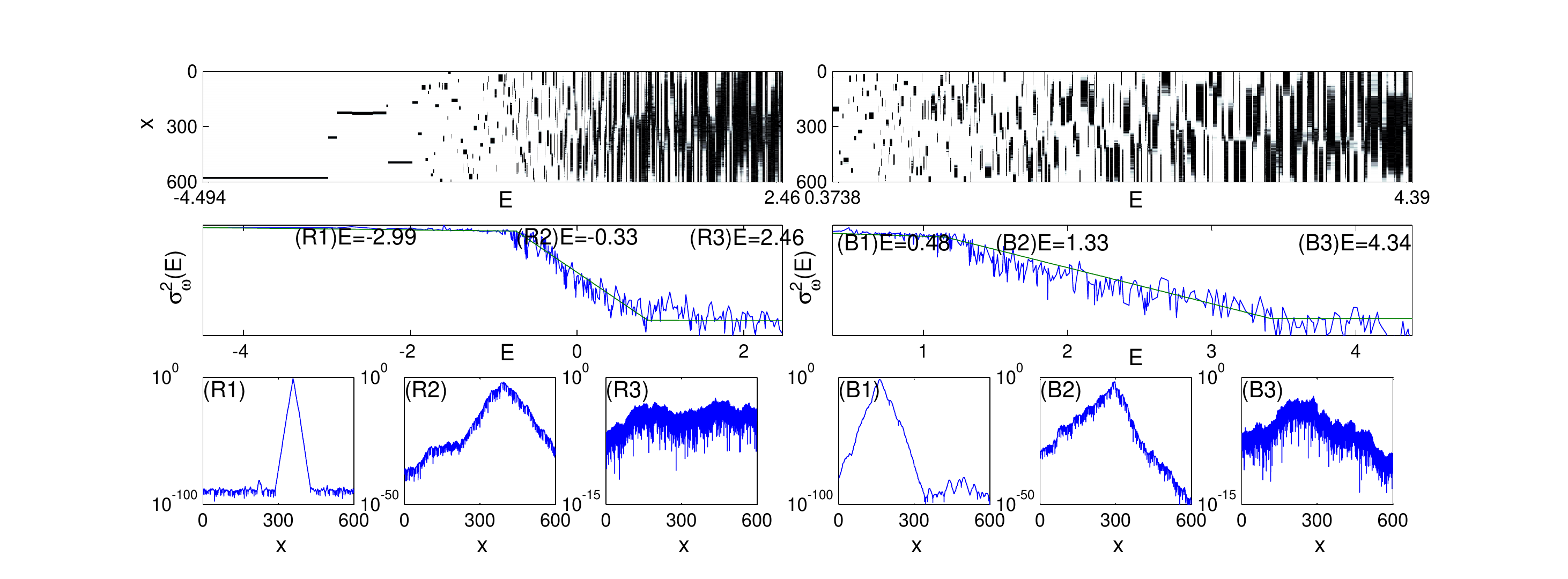} 
\caption{Left panels: $(top)$ example ($s=-1$) of density plot for the spectrum of a red detuned speckle potential. $(center)$ Comparison between the eigenstates $\psi_E$ of a single realization of the speckle potential and $\sigma^2_{\omega}\left[\psi_E\right]$. This will be discussed in \S \ref{subparso}. $(bottom)$ Density profiles of three eigenstates are plotted in a convenient logscale, in order to show the differences amongst the localized, delocalizing and delocalized  wave functions. Right panels: this figure is analogous to Fig. \ref{figexample_sigmaomega_r} (left panels) for a blue detuned speckle potential. In this case we have considered $s=1$.} 
\label{figexample_sigmaomega_r}
\label{figexample_sigmaomega_b}
\end{figure}

Let us now consider the behaviour of the excited states as a function of their energy $E$.
We expect the lower part of the spectrum to be affected mainly by the presence of the ``disordered'' potential, whereas at high energy the finite-size effect (the effect of the boundary conditions) should be predominant.
As an example of the behaviours discussed in this paragraph, in Fig. \ref{figexample_sigmaomega_r} (top panels) we show a density plot of the eigenstates, for a single realization of the speckle potential, for both attractive and repulsive cases (the functional $\sigma^2_{\omega}$ and the lower panels are discussed in \S \ref{subparso}). We observe the presence of three regions: localized states for low energy, extended states for high energy, separated by a wide intermediate region.
In the following we analyze the localization properties of the eigenstates as a function of their energy $E$, by considering their multifractal behaviour and localization length.
 
\subsection{Effective mobility edge in a box $(L\rightarrow\infty)$}
\label{me_intro}
In \cite{sanchezpalencia2007} it is shown that, under the Born approximation, there is a high-momentum cutoff $k_C$ such that all the matter waves with momentum $k>k_C$ are extended on every length scale (i.e. $\ell_{LOC}^{-1}(k\geq k_C)=0$). 

\begin{figure}[h]
\includegraphics[width=0.99\columnwidth]{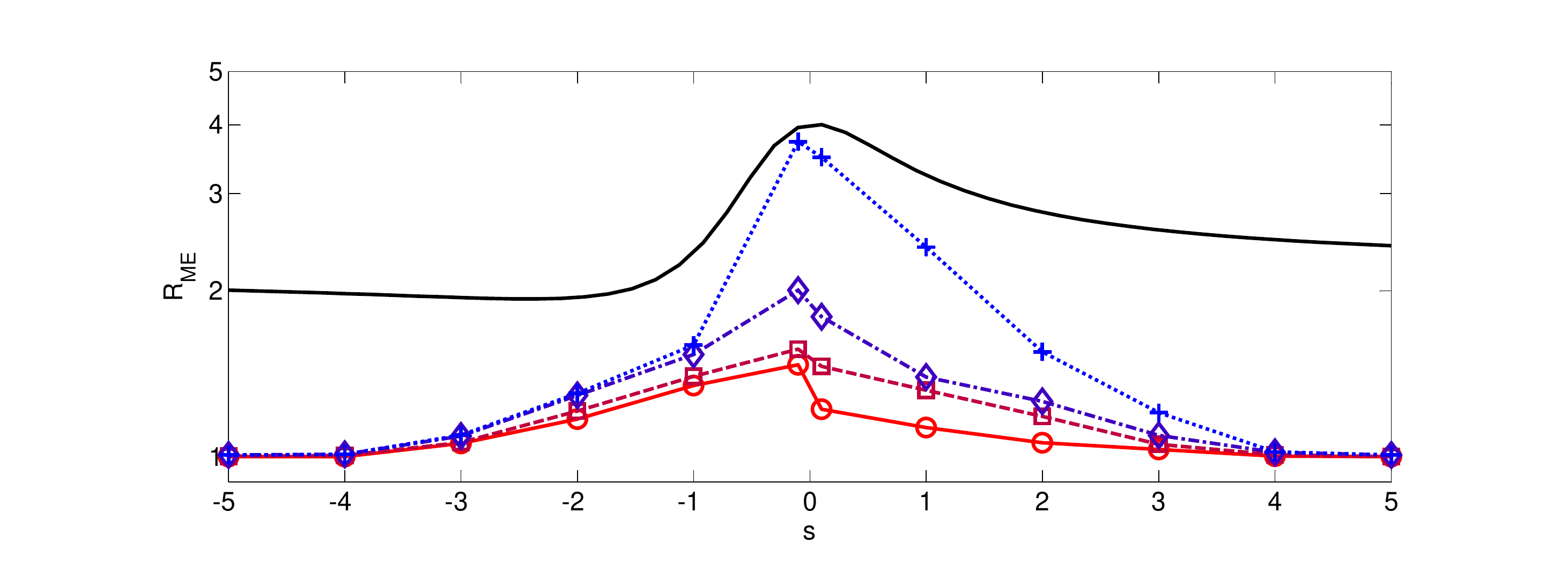} 
\caption{Mobility Edge Ratio  $R_{ME}\left(\epsilon=0.1, s, L\right)$ averaged on $50$ speckle realizations, with $L$ ranging from $4000$ (red solid line, ``$O$'' markers) to $7000$ (blue dotted line, ``$+$'' markers) and $\vert s \vert\in\left[0.1, 5\right]$. The solid black line is obtained from the $4^{th}$ order perturbation theory for $L=\infty$.}
\label{fig_ME_transition}
\end{figure}

\begin{figure}[h]
\includegraphics[width=0.99\columnwidth]{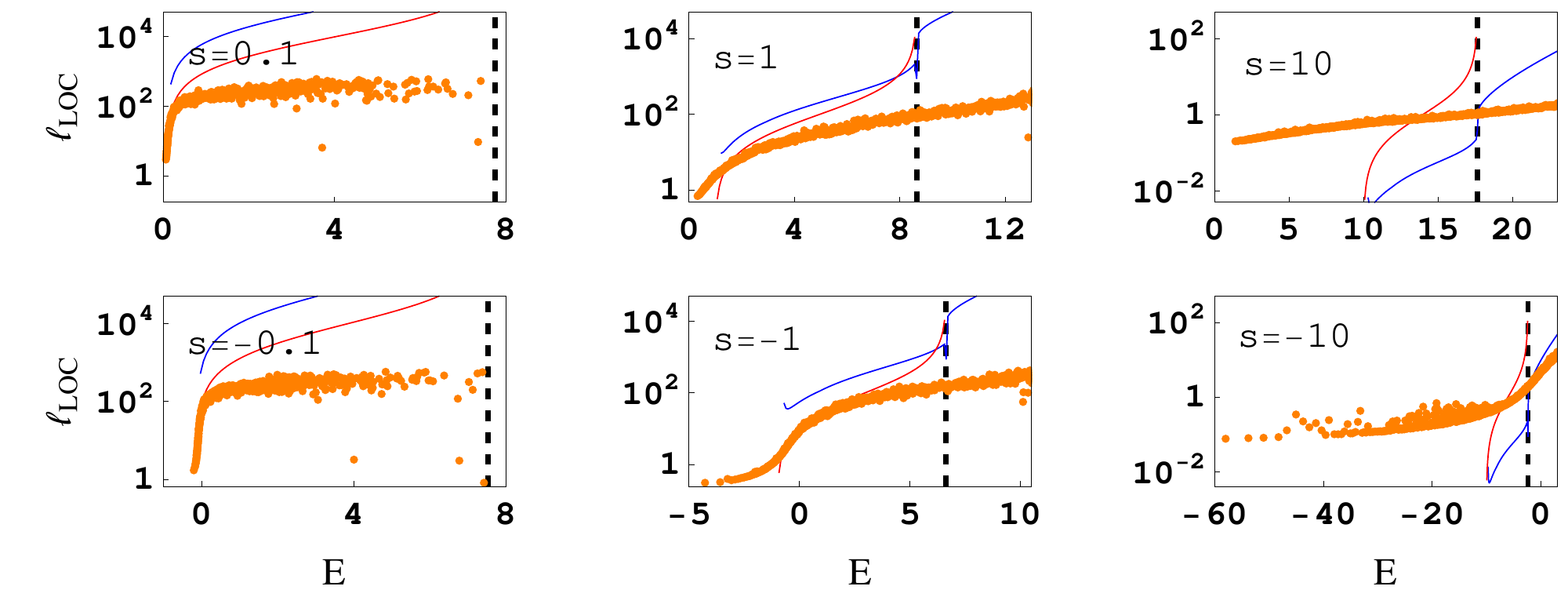} 
\caption{$\ell_{LOC}(E)$ for $-10\leq s\leq10$ and $L=800$. The figure shows a comparison between the simulations (orange dots, averaged over 50 realizations of the potential) and the theoretical results for $L=+\infty$ (second order - red line, fourth order - blue line).}
\label{fig_llocs_xp_vs_th}
\end{figure}

In our case (see also \cite{falco2010,modugno2006}) the expression for $\ell_{LOC}$ given in \cite{sanchezpalencia2007} reads (see \ref{MEcalculation})
\begin{equation}
\label{lloc_sE}
\ell_{loc}(s,E)=\begin{cases}
\displaystyle\frac{7.04}{s^2}\frac{E-s}{1-(0.88\pi)^{-1}\sqrt{E-s}} & E<E_C\\
+\infty & E\geq E_C
\end{cases}
\end{equation}  
under the condition 
\begin{equation}
\label{ournewbornassumption}
\vert s\vert\ll s_C=2(0.88\pi)^2\,.
\end{equation}
 In the limit of $\vert s\vert\rightarrow 0$ and $L\rightarrow\infty$, $\ell_{LOC}(E,s,L)$ diverges  in eq. (\ref{lloc_sE}) at $E\geq E_C$, where
\begin{equation}
\label{eqec}
E_C=s+(0.88\pi)^2.
\end{equation}
 However, the perturbative approach developed in \cite{lugan2009} shows that $\ell_{LOC}(E\geq E_C,s,L)$ is finite for $\vert s\vert>0$, though it increases significantly in a small interval around $E=E_C$. This behaviour is known as \textit{effective mobility edge}. Using the perturbative approach at the fourth order for infinitely extended systems, as explicitly computed in \cite{lugan2009}, we want to verify the presence of this effect also for the disorder in a box discussed in this work. To do so, let us define the ratio
\begin{equation}
R_{ME}(\epsilon,s,L)=\frac{\ell_{LOC}\left(E_C(s)+\epsilon,s,L\right)}{\ell_{LOC}\left(E_C(s)-\epsilon,s,L\right)}
\end{equation}
In fig. \ref{fig_ME_transition}, a comparison between the theoretical prediction for $L=\infty$ and our simulations for $L\in[4000, 7000]$ shows that also the finite size systems exhibit an effective mobility edge, but the growth rate of $\ell_{LOC}$ around $E=E_C$ decreases while reducing $L$. As plotted in fig. \ref{fig_llocs_xp_vs_th}, at $L\lesssim 1000$, the effective mobility edge is completely replaced by another regime, even at $s$ values satisfying the Born approximation. In the following we want to describe this regime, building a simple functional form that fits the numerical evidence at small $L$ values.

\begin{figure}[h]
\includegraphics[width=0.9\columnwidth]{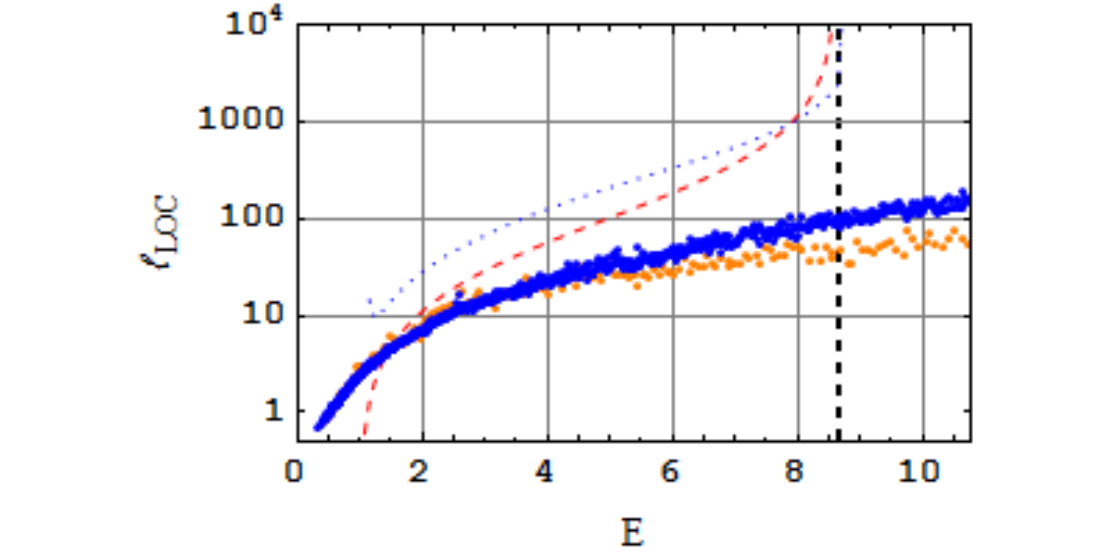} 
\caption{$\ell_{LOC}(E)$ for $s=1$ and $L=200$ (orange dots), $L=900$ (blue dots). The figure shows a comparison between the numerical simulations (dots, averaged over 50 realizations of the potential) and the theoretical prediction (second order - red dashed line, fourth order - blue dotted line). The dashed line marks $E_C$.}
\label{fig_llocL_xp_vs_th}
\end{figure}

\begin{figure}[h]
\includegraphics[width=0.9\columnwidth]{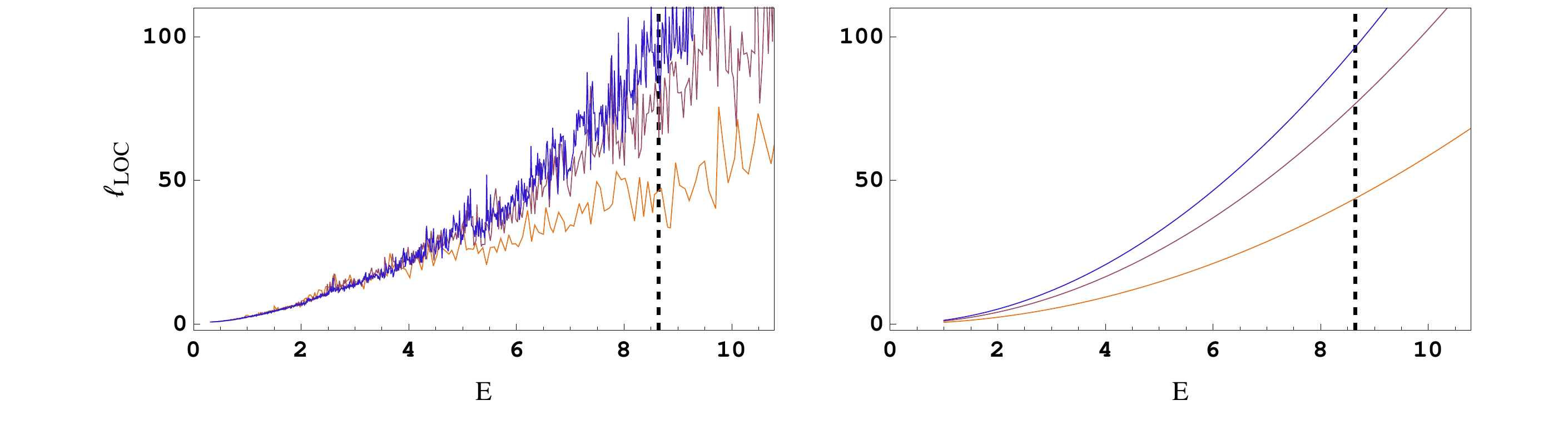} 
\caption{$\ell_{LOC}(E)$ for $s=1$ and $L$ between $200$ (orange) and $900$ (blue). The top figure shows the numerical dependence on $L$ while increasing $E$. $\ell_{LOC}$ does not show any sensibility to the ME (dashed line). The numerical results are averaged over several realizations of the potential. The bottom figure shows a fit of the same output, using the function $\hat{\ell}_{LOC}(E,L)=\exp\{g_1(L)\ln(E)+g_2(L)\}$ (chosen on the basis of the numerical evidence). The average value of the $R^2$  of the fit is $0.985$ across the fitted functions.}
\label{fig_llocL_xp_fit}
\end{figure}

\begin{figure}[h]
\includegraphics[width=0.95\columnwidth]{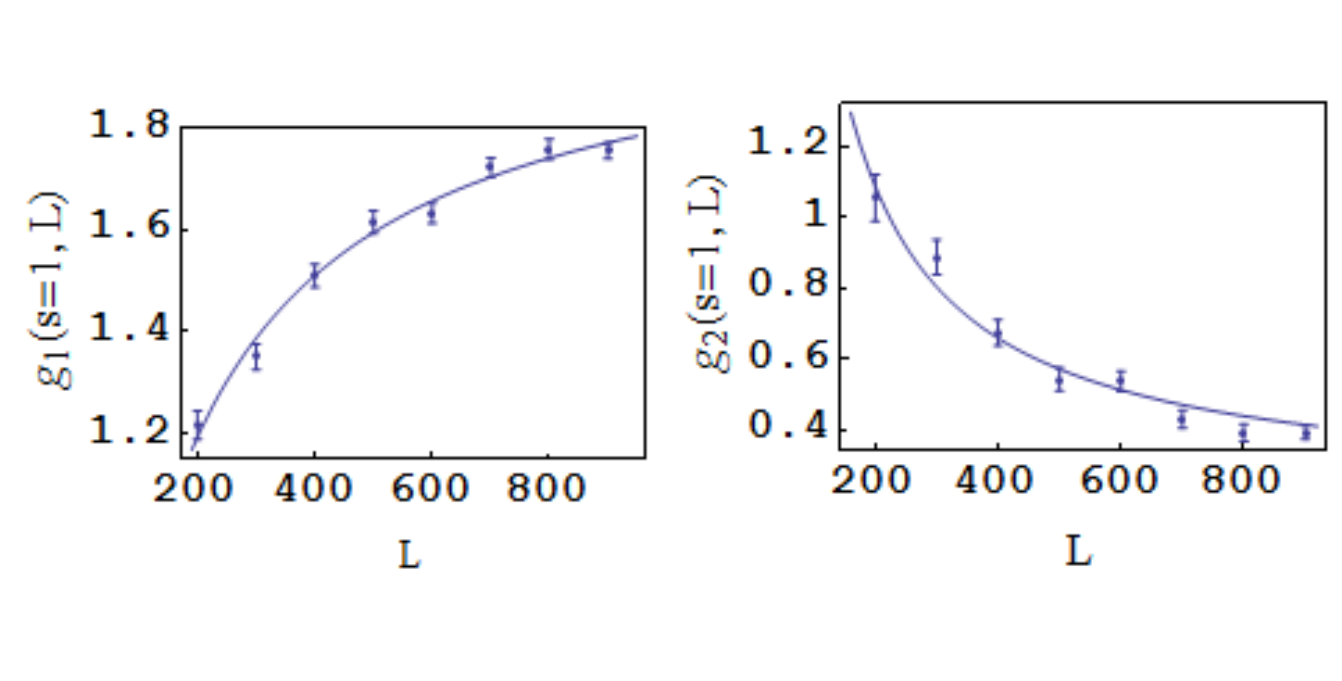} 
\caption{The figure shows the fitted values of $g_1$ and $g_2$ (dots) across the tested values of $L$, plotted together their respective fit errors (bars), considering $s=1$ as an example. The numerical results are fitted using the functions described in equations (\ref{lloc_sblu_1}) and (\ref{lloc_sblu_2}).}
\label{fig_lloc_L_param}
\end{figure}

As shown in fig. \ref{fig_ME_transition} and \ref{fig_llocL_xp_vs_th}, $\ell_{LOC}$ exhibits a dependence on $L$ both when the effective mobility edge regime is absent ($L\lesssim 1000$) and when it is being reached ($4000\lesssim L\rightarrow\infty$). However, this dependence asymptotically disappears at increasing $L$. Moreover, we verify that  $\ell_{LOC}(E< 0,s,L)$ looks like a straight line when plotted in a log$-$log scale as a function of $E$. This fact holds for all tested values of $s$ ($\vert s\vert\leq 10$) and $L\lesssim 4000$. These two elements allow us to define an empirical expression for $\ell_{LOC}(E>0,s,L\lesssim 1000)$:
\begin{equation}
\label{lloc_sblu_1}
\ell_{LOC}(E)\vert_{ s, L} = \exp\{g_1(s,L)\ln(E)+g_2(s,L)\}
\end{equation}  
We look for the functions of $L$ which fit $g_1$ and $g_2$ better. For all cosidered values of $s$ the following form fits the numerical data with a very good degree of approximation.
\begin{eqnarray}
\label{lloc_sblu_2}
\hat{g}_1(L)&=&\alpha_1(s)L/(\beta_1(s)+L)\\ 
\hat{g}_2(L)&=&\alpha_2(s)/(\beta_2(s)+L)+\gamma_2(s)\nonumber 
\end{eqnarray}  
Eq. (\ref{lloc_sblu_2}) implies the asympotic independence of $\ell_{LOC}$ from $L$ in the limit $L\rightarrow\infty$. Fig. \ref{fig_llocL_xp_fit} and \ref{fig_lloc_L_param} show an example of the overlap between Eq. (\ref{lloc_sblu_1},\ref{lloc_sblu_2}) and our numerical results .


\subsection{Self similarity of the eigenstates as a marker of the delocalization when lacking effective mobility edge $(L\lesssim 10^3)$}
\label{subparso}
For $(D>2)-$dimensional systems, there is an Anderson transition from localized to extended states \cite{evers2008}. The critical eigenstates show the multifractal behaviour associated with the transition also if the system is finitely extended \cite{aoki1983}. The 1D case has no Anderson transition at all (see \S~\ref{intro}), although multifractality has been observed in certain systems \cite{schreiber1991}.
  
Looking for a quantity able to distinguish among different degrees of localization, we compare the results obtained with $\sigma^2_{\omega}\left[\psi_{E}\right]$, defined in equation \ref{sodef}, against the other main observables defined in \S\ref{intrononint} and the variance of the density $\sigma^2\left[\psi_{E}\right]$. Despite no one eigenstate of our system is a multifractal, this functional turns out to be the only one able to react at the crossover between localized and extended states. We derive $\sigma^2_{\omega}\left[\psi_{E}\right]$ from the basic elements of multifractal analysis in \ref{app_sodef}. 

\begin{figure}[h]
\includegraphics[width=0.8\columnwidth]{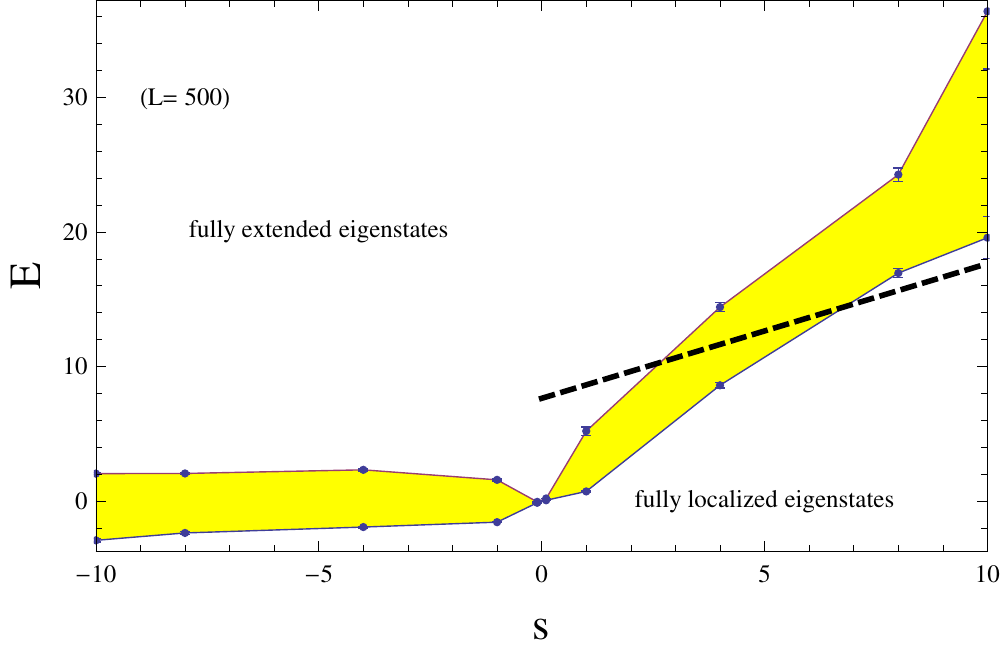} 
\caption{The amplitude of the intermediate region as a function of $s$. $E^{(a)}_C$ and $E^{(d)}_C$ are measured averaging the values obtained from eq. (\ref{eqso_fit}) over 50 realizations. The dashed line plots the predicted threshold for the Effective Mobility Edge.}
\label{figexample_sigmaomega3}
\end{figure}

The link between the possible density profile of the wavefunctions and the correspondent values assumed by  $\sigma^2_{\omega}\left[\psi_E\right]$ is shown in Fig. \ref{figexample_sigmaomega_r}.
For both localized and extended states, $\sigma^2_{\omega}\left[\psi_E\right]$ oscillates randomly around a constant value. In the intermediate delocalization region of the spectrum, $\sigma^2_{\omega}\left[\psi_E\right]$ decreases while $E$ is growing. In order to measure $E^{(a)}_C$ and $E^{(d)}_C$ (respectively the attachment point and the detachment point of the intermediate energy tranche where the delocalization process occurs), we fit $\ln\sigma^2_{\omega}\left[\psi_E\right]$ with a polygonal chain (see \ref{appnum} for further details).
The behaviour of $E^{(a)}_C$ and $E^{(d)}_C$ as a function of $s$ is shown in Fig. \ref{figexample_sigmaomega3}. We also check that the extent of the intermediate region is not affected by the system size $L$ for small enough systems ($L\lesssim10^3$). 

Fig. \ref{figexample_sigmaomega3} shows the localizing behaviour described above as a function of $s$: 
\begin{itemize}
 \item For $s>0$, the intermediate region is shifted towards higher energy levels  and its amplitude increases, for increasing values of $s$. This fact can be related to the distribution of the maxima of the potential, that are higher and more dispersed for larger values of $s$. Conversely, for $s<0$ the intermediate region is far more stable in mean and amplitude, according with a distribution of the maxima less sensibles to the $s$ variations and more concentrate around its mean value.
 
\item The delocalization process at increasing energy levels is not related to the effective mobility edge. In Fig. \ref{figexample_sigmaomega3} we consider the length scale discussed in \S \ref{me_intro} for which we have verified the absence of an effective mobility edge. The figure shows that in this case $E_C(s)$ defined in eq. (\ref{eqec}) is not in agreement with the position of the \textit{intermediate} region $[E_C^{(a)},E_C^{(b)}]$. This is a non trivial behavior, because it demonstrates that the the boundaries of the system can not only stop the exponential decay of the eigenstates density profile, but also reduce the localization length, in accordance with \S \ref{me_intro}. 
\end{itemize}

Since the intermediate region separates Anderson localized states from extended states, weak localization phenomena are likely to be observed in this energy band.

\section{Summary}
\label{secend}
In this work we have investigated the density profile of the single particle eigenfunctions in a \textit{box-plus-speckle-potential} system. We have analyzed how the parameters $s$, $L$ and $E$ influence the density profile of the eigenfunctions, giving special attention to the degree of localization. 

The ground state exhibits a very weak dependence on $L$. For increasing $\vert s \vert$, a smooth passage is observed from Anderson localized states to eigenstates bounded in a single well (or couple of barriers). However, it is possible to partition the localization process into two distinct regimes. The localization as a function of $\vert s\vert$ is faster for the attractive case.

For low energy eigenstates, the localization point and so the center of mass can shift abruptly while tuning $\vert s\vert$. This is due to an occasional degeneracy with the next energy level. We have linked the probability of observing this shift to the eigenstates participation ratio and to the system finite extension - the latter contribution being relevant only for $\vert s\vert\ll 1$.

For $L$ small enough ($L\lesssim 1000$), the system does not exhibit any effective mobility edge while delocalizing at increasing $E$, even when being still localized $E\geq E_C$. This is due to the finite size $L$ and how it affects the long range self-correlation of the system. In this regime, we have described the behaviour of the localization length as a function of $s$, $L$ and $E>0$. For increasing energy, we have also found that the eigenstates of this system can be divided in three distinct  regions according to their degree of localization. The quantity developed to observe this partition may find an application also in other contexts as an indicator of the degree of self-similarity in data series. 

For $L$ big enough, we have observed an approach to the effective mobility edge, consistently with the theory for $L\rightarrow\infty$.

\section{Acknowledgements}
The author wants to thank M. Modugno for the useful discussions and advices during the whole developement of this work. 
Useful comments by L. Sanchez-Palencia and B. Shapiro are acknowledged.

\appendix
\section{The effective mobility edge for infinitely extended systems expressed in our units}
\label{MEcalculation}
As stated in \cite{sanchezpalencia2007}, the Born approximation in this context assumes the inequality
\begin{equation}
\label{bornassumption}
V_0\tilde{\xi} \ll (\hbar^2k/m)(k\tilde{\xi})^{1/2},
\end{equation} 
where $\xi=0.88\pi\tilde{\xi}$. Using in Eq. (\ref{bornassumption}) the definition of $s$ given in Section \ref{intrononint}, we obtain the following equivalent condition:
\begin{equation}
\label{ourbornassumption}
E \gg s + Cs^{4/3},
\end{equation}
where $C=2^{-4/3}(0.88\pi)^{-2/3}\simeq 0.2$. Under this condition, we have
\begin{equation}
\label{mobilityedge}
\ell_{LOC}^{-1}(k)=\frac{\pi m^2V_0^2\tilde{\xi}}{2 \hbar^4 k^2}f(k) 
\end{equation} 
where $\sqrt{\pi/2}f(k)= \sqrt{\pi/2}(1-k\tilde{\xi})\Theta(1-k\tilde{\xi})$ is the power spectrum of the potential in case of an infinitely extended system. Substituting the definition of $s$ again, Eq. (\ref{mobilityedge}) in our framework becomes
\begin{equation}
\label{ourmobilityedge}
\ell_{LOC}^{-1}(s,E)=\frac{s^2}{8\cdot0.88}\frac{1-\frac{\sqrt{E-s}}{0.88\pi}}{E-s}\Theta(s+(0.88\pi)^2-E)
\end{equation}
which leads to eq. (\ref{lloc_sE}). From eq. (\ref{ourbornassumption}) and (\ref{ourmobilityedge}), it follows that
\begin{equation}
\label{ournewcond}
s + Cs^{4/3}\ll E< s+(0.88\pi)^2,
\end{equation}
that leads to a stronger requirement than (\ref{bornassumption}) for allowing the good definition of $\ell_{LOC}$, as stated in eq. (\ref{ournewbornassumption}).

\section{CLP effect}
\label{app_clp}
In the following we are going to prove that the result in equation (\ref{eqclpdistr}) holds except for a small interval around $s=0$. In this section we use the notation that ``$a\sim b$'' means ``$b$ is the probability density function of the random variable $a$''.  Naively speaking, a change of localization point event (CLP event in the following) occurs when two next localized eigenstates exchange their positions in the spectrum because of a variation of $s$. Without loss of generality, let us consider the ground state $\psi_0$ and the first excited state $\psi_1$, with their respective energy eigenvalues $E_0$ and $E_1$. So we have
\begin{equation}
\label{clp_cond}
p_{CLP}\left(s\right)=p\left(E_0=E_1\vert s\right).
\end{equation}
As shown before, low energy eigenstates of the discussed model are well described by 
\begin{equation}
\label{psilloc}
\psi_i(x)=c\left[e^{-\frac{\vert x-x_{LOC}\vert}{\ell_{LOC}}}1_{x\in D_{tails}} + g(x)1_{x\notin D_{tails}}\right]
\end{equation}
where $D_{tails}$ is the domain where the wavefunction exponentially decays and $g(x)$ is the ground noise out from  $D_{tails}$. We want to explicit the value of $E_i$, in order to solve the equation (\ref{clp_cond}). We can state that
\begin{equation}
\label{eiexplicit}
E_i = K_i+sV_i
\end{equation}
using the fact that the Hamiltonian is separable. For the kinetic term, approximating $g(x)\simeq0\simeq e^{\frac{\vert x-x_{LOC}\vert}{\ell_{LOC}}}$, we have
\begin{eqnarray}
\label{eqk}
K_i&=&\int_0^L\psi_i(x)\frac{\partial^2}{\partial x^2}\psi_i(x)dx\nonumber\\
&\simeq&\frac{c^2}{\ell_{LOC}^2}\int_0^Le^{-2\frac{\vert x-x_{LOC}\vert}{\ell_{LOC}}}dx\\
&\simeq&\frac{1}{\ell_{LOC}^2}\nonumber
\end{eqnarray}
this approximation holds for $\vert s\vert$ large enough that boundary effects are negligible.  
For the potential term we have
\begin{equation}
V_i =\int_0^L\vert\psi_i(x)\vert^2v(x)dx\simeq \sum_j \psi^2_{ij}v_j
\end{equation}
where we approximate the continuos system applying an arbitrary discretization. Neglecting the self correlation of the potential, we can consider all the $v_j$ as i.i.d. stochastic variables. Since $v_j$ follows a negative exponential distribution, we have 
\begin{equation}
\psi^2_{ij}v_j\sim\Theta\left(v_j\right)\psi^{-2}_{ij} e^{-\psi^{-2}_{ij}v_j}.
\end{equation}
A sum of independent exponentially distributed variables with different mean life values fits the generalized Erlang distribution. So we are able to determine its variance in a closed form. Let us partition the system in $\xi$ long intervals in order to guarantee the indipendency of the random variables. 
\begin{equation}
\sigma^2\left[V_i\right]\simeq \sum_j \psi^4_{ij}\simeq  \frac{L}{\xi}PR(s,L)^{-1}
\end{equation}
On average, it holds that $\sigma^2\left[V_0\right]=\sigma^2\left[V_1\right]$ and $\mu\left[V_0\right]=\mu\left[V_1\right]$. Approximating the distributions of $V_0$ and $V_1$ with two normal distributions, we have
\begin{equation}
\label{eqdv}
\frac{E_0-E_1}{s} = V_0-V_1:=\tilde{v}\sim \sqrt{\frac{L}{\xi}\frac{PR(s,L)}{2\pi}}e^{-\frac{ L PR(s,L) \tilde{v}^2}{2\xi}}
\end{equation}
Replacing equations (\ref{eiexplicit}), (\ref{eqk}) and (\ref{eqdv}) into equation (\ref{clp_cond}), we obtain
\begin{equation}
p_{CLP}(s,L) = p(\tilde{v}=0\vert s,L) = \sqrt{\frac{L}{\xi}\frac{PR(s,L)}{2\pi}}
\end{equation}
that leads to equation (\ref{eqclpdistr}). 

\section{Definition of $\sigma_{\omega}^2$}
\label{app_sodef}
Generally speaking, a scale invariance property of a given function $f(x)$ can be numerically verified partitioning it into clusters of increasing size. The necessary condition that $f(x)$ must satisfy for being considered as a fractal (or a multifractal) is the independence of a certain  quantity $\omega\left(q,\Lambda\right)$ - defined later on - from the chosen size $\Lambda$ of the clusters. In the following we specialize $f(x)=\vert\psi(x)\vert^2$. Let us introduce the measures
\begin{equation}
\label{measureinpar_v2}
\mu_k(\Lambda)=\int_{k\Lambda}^{(k+1)\Lambda}dx f(x)\quad k=0,\dots,k_{max}(\Lambda)
\end{equation}
where the interval $[0,\left(k_{max}+1\right)\Lambda\equiv L]$ is the spatial domain of the system for every $k_{max}\in\mathbb{N}$. For each $q\in\mathbb{Z}$ and for each acceptable value of $\Lambda$, we can compute the $q-$th moment $P_q(\Lambda)$:
\begin{equation}
\label{momentinpar_v2}
P_q(\Lambda) = \sum_{k}\left\lbrack\mu_k(\Lambda)\right\rbrack^q, \quad q\in\mathbb{Z}.
\end{equation}
For a self-similar function, it holds that
\begin{equation}
\label{momentinpar_trend}
P_q(\Lambda)\propto \Lambda^{\omega(q)}.
\end{equation}
When the relation (\ref{momentinpar_trend}) is verified, the shape of $\omega(q)$ defines the type of self-similarity of the function (if $\omega(q)$ is linear with a single slope, $f(x)$ is fractal; if $\omega(q)$ is linear with two different slopes for positive and negative $q$, $f(x)$ is multifractal; otherwise $f(x)$ is not a fractal in any sense). For every $f(x)$ (even for the non self similar ones), the following quantity is well defined:
\begin{equation} 
\omega(q,\Lambda) = \ln P_q(\Lambda)/\ln \Lambda.
\end{equation} 
Given that $f(x)$ can be a (multi)fractal \textit{only if} the numerical dependence of $\omega(q,\Lambda)$ from $\Lambda$ is almost absent, we can introduce a quantity to measure \textit{how much far} is $f(x)$ from being self-similar in some sense:
\begin{equation}
\label{eqso}
\sigma^2_{\omega}[f] = \sum_q\left\langle\omega^2(q,\Lambda)-\left\langle\omega(q,\Lambda)\right\rangle_\Lambda^2\right\rangle_\Lambda.
\end{equation} 
In a $(D>2)-$speckle system, the presence of an Anderson transition at $E=E_{AT}$ should lead to 
\begin{equation}
\label{eqlimso}
\lim_{E\rightarrow E_{AT}}\sigma^2_{\omega}\left[\vert\psi_E\vert^2\right]=0
\end{equation}

\section{Numerical methods and criteria}
\label{appnum}
In the following section we summarize some methods that we use to obtain the results contained in Sec. \ref{secgs} and \ref{seces}. 

\textit{Generation of the eigenstates}. We use the following prescription:
${H}_{ij} = {1}/{\Delta\tilde{x}^2}\left(2\delta_{ij}%
-\delta_{i+1,j}-\delta_{i-1,j}\right) + s \tilde{I}(j\Delta \tilde{x})%
\delta_{ij}$, with vanishing boundary conditions, $\psi_0=\psi_{N+1}=0$.
The number of grid points $N$ is chosen so that every fluctuation of the potential is sampled with at least $8$ points (i.e.: $N\geq8L/\xi$).

\textit{Identification and quality check of the exponential tails}. We need to identify four $x$ values $(x_1, \dots, x_4$ in ascending order$)$ in order to select  the two tails properly.  The routine starts choosing the points $x_2$ and $x_3$, near the localization point $x_{LOC}=\textrm{argmax}\left[\vert\psi\left( x\right) \vert\right] $ but far enough to exclude the local fluctuation of the potential around which the eigenfunction is localized (i.e.: $x_{2,3}=x_{LOC}\pm\xi$). \newline
The terminal point of the right tail $x_4$ is found going through the $\vert\psi(x)\vert$ values for increasing $x>x_3$, until $\vert\psi(x)\vert\leq\psi_{noise}$ or $x\geq L-\xi$. The routine identifies $x_1$ using a symmetrical rule to the one described above. The threshold $\psi_{noise}$ has the form
\begin{equation}
\psi_{noise}(\alpha) := e^{\left[\alpha\max(\ln\vert\psi(x)\vert)+(1-\alpha) \min(\ln\vert\psi(x)\vert)\right]}\nonumber
\end{equation}
where $\alpha$ is a user defined parameter.\newline
We checked the exponential shape of the tails systematically, computing the $R^2$ for the fit used to measure $\ell_{LOC}$. When $R^2$ value return an ambiguous indication, we verify the shape of the tails by a direct observation.

\textit{Check for the physical meaning of the ground noise}. 
In \S~\ref{intrononint} we introduce the localization volume $D_{LOC}$ to better understand the 
local influence of the boundaries over the density profile. $D_{LOC}$ is well defined only if the 
abrupt interruption of the exponential decay of the eigenfunctions out of the interval $\left
[x_1,x_4\right]$ is not a numerical effect - at least in the discussed region of $s, L$. We 
have verified this fact performing the three tests described below over a sample of speckle 
realizations: 
$(i)$ computation of the same eigenstate using different computers; 
$(ii)$ computation of the same eigenstate using a re-sampled version of the speckle potential 
with a different $N$ (produced by a spline interpolation); 
$(iii)$ comparison of the stability of $\psi(x)$ to an ``adjusted version'' of the same 
eigenstate $\tilde{\psi}(x)$. The latter is obtained extrapolating the esponential decay outside  
$\left[x_1,x_4\right]$ until the machine double epsilon ($\sim 5\cdot 10^{-324}$ in this case) or 
a boundary of the system is reached. This test has been performed applying the following condition to check the considered eigenstates
\begin{equation}
\left\vert\left[\mathbf{1}-\hat{n}\circ\hat{\mathcal{H}}\right]\psi_{E}\right\vert^2<\left\vert
\left[\mathbf{1}-\hat{n}\circ\hat{\mathcal{H}}\right]\tilde{\psi}_{E}\right\vert^2\nonumber
\end{equation}
where $\hat{n}\left[v\in\mathbb{R}^N\right]:=v/\vert v\vert$. Ideally it should hold that $\hat
{n}\circ\hat{\mathcal{H}}\psi_{E}=\hat{\mathcal{H}}\psi_{E}/E$, but we prefer the criterion 
above to avoid the error over the estimate of the eigenvalue $E$.

\textit{Measure of $E^{(a)}_C$ and $E^{(d)}_C$ in \S \ref{subparso}}. 
The results shown in Fig. \ref{figexample_sigmaomega_r}, \ref{figexample_sigmaomega_b} and \ref{figexample_sigmaomega3} are obtained fitting $\ln\left[\sigma^2_{\omega}\right](E)$ with a polygonal chain. More in detail, we use the following function
\begin{eqnarray}
\label{eqso_fit}
\hat{\sigma}^2_{\omega}\left[\psi_E\right]&=&\exp\lbrace \left(c_1E+c_2\right)\Theta(E^{(a)}_C-E)\Theta(E-E_{gs})\nonumber\\
&+&\left(c_3E+c_4\right)\Theta(E-E^{(a)}_C)\Theta(E^{(d)}_C-E)\nonumber\\
&+&c_5\Theta(E-E^{(d)}_C)\rbrace\nonumber
\end{eqnarray} 
from which we obtain an estimate of $E^{(a)}_C$ and $E^{(d)}_C$, respectively the attachment and the detachment point of the crossover region.


\end{document}